# Different degrees of complexity in multiparticle production at the LHC energies with the transition from soft to hard processes

Hirak K. Koley,[1] Arindam Mondal,[2] Somnath Kar,[1] Sreejita Mukherjee,[1] Joyati Mondal,[1] Argha Deb[1,3] and Mitali Mondal[1,3]

[1] Nuclear and Particle Physics Research Centre, Department of Physics, Jadavpur University, Raja S. C. Mallick Road, Kolkata - 700032, West Bengal, India.
[2] RCC Institute of Information Technology, Beliaghata, Kolkata - 700015, West Bengal, India
[3] School of Studies in Environmental Radiation & Archaeological Sciences, Jadavpur University, Raja S. C. Mallick Road, Kolkata - 700032, West Bengal, India.

Correspondence should be addressed to Arindam Mondal; arindammond@gmail.com
Somnath Kar; somnathkar11@gmail.com
Mitali Mondal; mitalimon@gmail.com

## Abstract

We apply a complex network-based method of visibility graph to explore the degree of complexity and fractal nature in the multiparticle production process at the LHC energy regime. For the investigation, we have used proton-proton (pp) collision events at $\sqrt{s}$ = 7 and 13 TeV and proton-nucleus (p-A) collision events at $\sqrt{s_{NN}}$ = 5.02 TeV, generated using a hybrid Monte-Carlo model, EPOS3 with hydrodynamical evolution. Particle production in high energy collisions becomes gradually harder with the increase of the produced particle's transverse momentum. We have presented for the first time a detailed analysis of the change of fractal behaviour as the process becomes gradually harder than the soft ones for the above-mentioned colliding systems. The study reveals that the degree of complexity depends on the hardness of the particle production process. System size dependency, as well as energy dependency of fractal nature, are also exhibited by the present analysis.

## 1. Introduction

Development of complex network studies in the last decade allowed the characterization of many types of systems in nature that contain many components interacting with each other intricately. The method of complex network studies has recently been used in the analysis of a time series. Scientists around the globe have developed many methods to capture the geometrical structure of time series from complex network aspects such as cycle network [1], correlation network [2], visibility graph [3], recurrence network [4] and many others. We use the visibility graph method, which is a Deterministic Road Technique, for our present analysis. A visibility graph maps a time series into a network. This network inherits various properties of the time series and the study of the network reveals nontrivial information about the series itself. It has been reported that it can transform the periodic time series into regular graphs, random series corresponding to random graphs [3] and fractal series into scale-free graphs [3]. The special feature of the visibility graph method is that it can be applied to a finite time series or time series like data. In reality, a time series is always finite. Therefore, visibility graph analysis has a wide range of practical applications. For example, the visibility graph technique has been applied in searching for the hidden geometry of traffic jamming [5], in studying energy dissipation rates in three-dimensional fully developed turbulence [6], in analysing exchange rate series [7], in searching for fluctuation and geometrical structure of magnetization time series of two-dimensional Ising model around critical point [8], in investigating human heartbeat dynamics [9–11] and in searching for fluctuation in multiparticle production in high energy collisions [12–21]. S. Bhaduri et al. [22] used the visibility graph analysis method to look for the signature of phase transition in the Pb-Pb collision data sample at 2.76 TeV per nucleon pair taken from CERN open data portal. They [23] have also applied symmetry scaling based complex network approach to study exotic resonance/hadronic states by analysing the Pb-Pb collision data sample at 2.76 TeV and following the same approach, they have studied pp collision data at 8 TeV from CMS Collaboration, taken from CERN open data portal. Zhao et al. have applied the visibility graph method to investigate the fluctuation and geometrical structures of magnetization time series. Lacasa et al. [24] have also shown how one can apply a classical method of complex network analysis to quantify the long-range dependence and fractality of a time series.
For the last thirty years, the Physics of Relativistic High Energy Collisions has been on the frontier of scientific research activities throughout the globe. According to the accepted theory, the tiny Universe of high energy density and temperature, immediately after the Big Bang, evolved through a state of an exotic phase of partonic matter, the Quark-Gluon Plasma (QGP) phase [25–34], that comprises quarks and gluons - the elementary particles. The formation of QGP in the laboratory was confirmed by experiments at the Relativistic Heavy Ion Collider (RHIC) at the Brookhaven National Laboratory [35–38] after the CERN's declaration of the sign of the formation of QGP- like-new state of matter at Super Proton Synchrotron (SPS) [39, 40]. Recently, ongoing experiments at the Large Hadron Collider (LHC) [41, 4] are studying heavy-ion collisions with heavier nuclei and at higher centre-of-mass energy per nucleon pair as compared to those in RHIC, to understand the properties of the QCD-plasma. In extracting the signals of the QGP state in heavy-ion collisions, the proton-proton (pp) collisions at the same energy serve as the baseline assuming the collisions being as elementary interactions. To establish the formation of a QGP like medium in heavy-ion collisions at the ultra-relativistic energy, one needs to disentangle the cold nuclear matter effects which may have aroused from the so-called cold nuclei before the collisions. The asymmetric collisions like proton-lead (p-Pb) play important role in this regard at the LHC. For a long time, the QGP was believed to be formed only in heavy-ion collisions and different hydrodynamical models remained successful in explaining the key features of the QGP. However, recent studies on hadron-hadron or hadron-on-ion collisions (small systems) at the RHIC and the LHC energies find unambiguous features which are very much similar to the heavy-ion collisions in terms of formation of QGP-like medium in small systems. Though the small collision systems exhibit a strong indication of anisotropic flow which resembles the heavy-ion data, it underestimates the key feature like the energy loss of high-$pT$ particles/jets, the so-called "jet-quenching". Therefore, the fascinating outcomes of small collision systems need to be investigated further with all sorts of available tools without any anticipation.

The study of large density fluctuations in multiparticle spectra in high energy interactions provides information regarding the dynamics of the particle production processes. Various analyses have revealed that pionization processes are self-similar which suggests the existence of fractal geometry of the particle production mechanism [43–48].

Taking motivation from the recent studies in explaining the fractal nature in multiparticle productions using the visibility graph technique in the high energy interactions [13, 19, 49, 50], we further dive into the search for the same in pp collisions at the centre-of-mass energy, $\sqrt{s}$ = 7 and 13 TeV and in p-Pb collisions at $\sqrt{s_{NN}}$ = 5.02 TeV, simulated by a $p$QCD inspired

Monte-Carlo model, EPOS3, using the method of visibility graph [3, 19]. We have mapped the event-by-event pseudorapidity ($\eta$) distribution of produced charged particles into a scale-free network to investigate the possible fractal nature in the collisions of high energy particles. The detailed comparison of hadron-hadron (pp) and hadron-on-ion (p-Pb) collisions has been illustrated in terms of the topological parameter obtained from the visibility graph technique.

The remainder part of this article is structured as follows. In Sec. 2, the method of visibility graph technique has been discussed. We have given a brief description of the EPOS3 model in Sec. 3, followed by the Result and Discussion in Sec. 4 and Conclusion in Sec. 5.

## 2. Visibility Graph Technique

The visibility graph algorithm maps time series $X$ to its visibility graph [3]. Figure 1 illustrates the basics of the method. Let us first consider that at any instant the $i^{\text{th}}$ point of the time series is $X_i$. Each data value is deliberated to be a node. Any two nodes are connected if they can see each other, namely, a straight visibility line exists between them. Formally, two arbitrary data values ($t_a$, $X_a$) and ($t_b$, $X_b$) are visible to each other if any other point ($t_c$, $X_c$) between them satisfies the criterion

$$X_c \leq X_b + (X_a - X_b)\frac{(t_b - t_c)}{(t_b - t_a)} \tag{1}$$

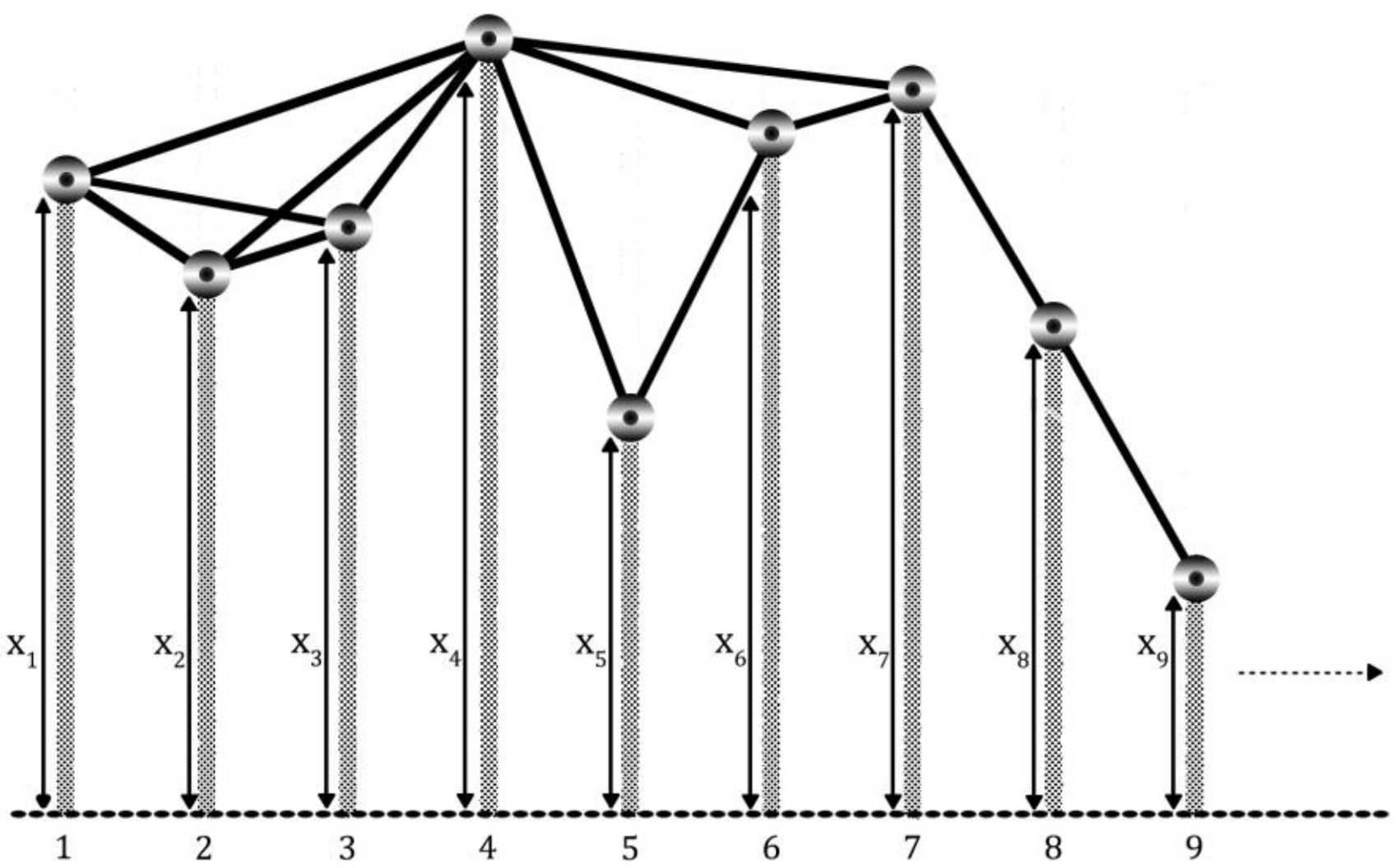

Fig. 1: Schematic representation of visibility graph for time series $X$.

From the graph theory, we know that the degree of a node is the number of connections or edges that the node has with other nodes. The degree distribution, say $P(k)$, of a network formed from the time series, is defined as the fraction of nodes with degree $k$ in the network. Thus, if there are $n$ number of nodes in total in a network and $n_k$ of them have degree $k$, then $P(k) = \frac{n_k}{n}$.

The scale-freeness property of the visibility graph states that the degree distribution of its nodes must satisfy a power-law behaviour as

$$P(k) \sim k^{-\lambda_p} \quad (2)$$

where $\lambda_p$ is a constant and it is known as Power of the Scale-freeness in Visibility Graph (PSVG). It manifests the degree of complexity and fractal nature of the time series and has a linear relationship with the dimension of the data/time series [3, 24, 51].

## 3. EPOS3 Monte-Carlo Simulation

EPOS3 [53] is a hybrid Monte-Carlo event generator based on the Gribov-Regge theory of multiparton scattering [58]. It has three major features: a flux-tube initial condition, 3+1D viscous hydrodynamics and a hadronic afterburner modelled via UrQMD model [54]. It uses a similar technique for particle production at the LHC energy in proton-proton (pp), proton-nucleus (p-A) and nucleus-nucleus (A-A) collisions.

In EPOS3, each binary interaction (elementary scattering of partons) is represented by a parton ladder [55]. There are two types of parton ladders: open (for inelastic collisions) and closed (for elastic collisions). In this model, beam remnants can possibly take part in the collision process enhancing the particle yield. Each parton ladder is considered as a longitudinal colour field or flux tube, carrying transverse momentum ($p_T$) of the hard scattering. The flux tubes expand and at some stage get fragmented into string segments of quark-antiquark pairs. In the case of high-multiplicity events, due to a large number of elementary parton-parton scatterings, many flux tubes are produced which lead to a highly dense medium of coloured string segments [56]. The string segments in the bulk matter, which do not have enough energy to escape, form a “core” of thermalized partons that undergoes hydrodynamical expansion following a 3 + 1D viscous hydrodynamic evolution followed by usual Cooper-Frye mechanisms of hadronization. After that, the hadronic evolution takes place till the freeze-out of the “soft” (low transverse momentum ($p_T$)) hadrons. On the other hand, the “hard” particles or the high-$p_T$ jet-hadrons, hadronized by Schwinger’s mechanism, originate from the less dense medium of high-energy string segments in the periphery of the bulk matter i.e the so-called “corona”. The “semi-hard” or the intermediate-$p_T$ particles originate from the string segments with enough energy to escape the bulk matter. These string segments, while escaping from the bulk matter, may pick up quark/antiquark from the medium. As a result, the intermediate-$p_T$ particle inherits the properties of the bulk medium [57].

Using the EPOS3 model, we have generated 10 million minimum-bias events for pp collisions at $\sqrt{s}$ = 7 & 13 TeV and p-Pb collisions at $\sqrt{s_{NN}}$ = 5.02 TeV. The EPOS3 model comes with options of switching ON or OFF the hydrodynamical evolution of particles from the core. In this study, we have only taken the events with the hydrodynamical evolution of produced particles (hereafter termed as “with hydro") only as it has been successful in explaining most of the LHC findings [55, 57–62]. In asymmetric collisions like proton-lead, the nucleon-nucleon centre-of-mass system is moving in the laboratory frame with a rapidity of $y_{NN}$ = -0.465 in the p-going direction. The convention yields that the laboratory frame rapidity, $y$ and the p-Pb centre-of-mass system rapidity, $y^*$, are related by $y^*$ = $y$ - 0.465. The pseudorapidity ($\eta$) variable also follows the same convention where the centre-of-mass frame is shifted from the laboratory frame as, $\eta_{cms}$ = $\eta_{lab}$- 0.465 [63]. In the remaining texts, we termed $\eta_{cms}$ as $\eta$ to make it simpler and cope with experimental notations.

To validate the EPOS3 generated events in p-Pb collisions used for the present study, we have compared the charged particle yields measured by ATLAS collaboration [64].

Figure 2 shows the invariant yields of charged particles in - 0.5 ≤ $y^*$ < 0.5 for EPOS3 generated with hydro events compared to ATLAS data in p-Pb collisions at $\sqrt{s_{NN}}$ = 5.02 TeV and we found that the model reproduces the data successfully except at very low $p_T$ region.
For the rest of the analysis, events are chosen with a minimum of three charged particles in the kinematic interval $p_T$ > 0.1 GeV/c and $\eta$ < 2.5 to include all charged particles produced in hard as well as soft processes.

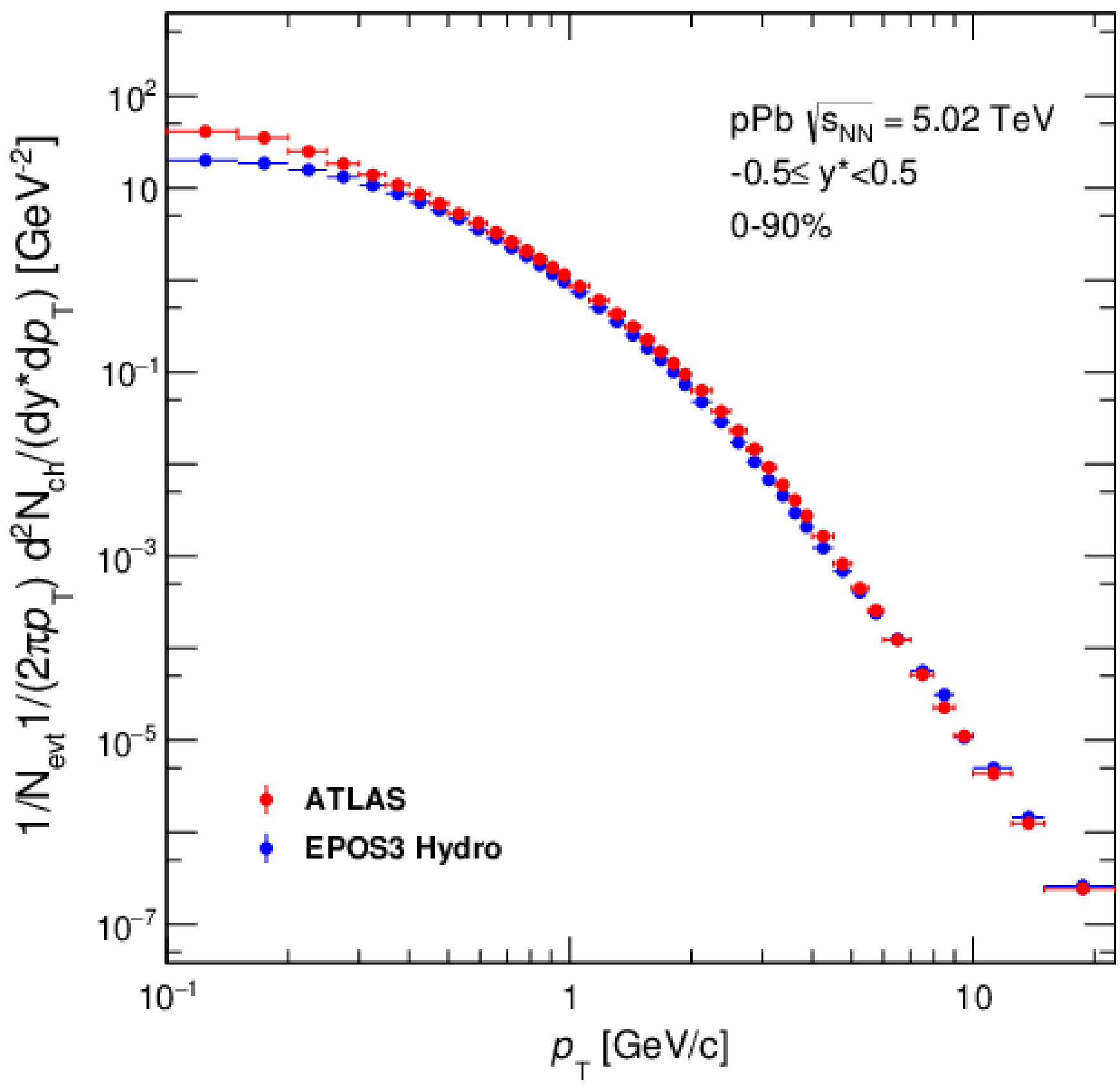

Fig. 2: (Color online) Invariant differential yields of charged particles as a function of transverse momentum from EPOS3 with hydro generated events for 0-90% centrality interval in p-Pb collisions at $\sqrt{s_{NN}}$ = 5.02 TeV in the rapidity interval ($-0.5 \leq y* < 0.5$), compared to ATLAS data.

## 4. Results and Discussion

Event-by-event fluctuations are being captured by converting the time series like data of single event pseudorapidity distribution into the corresponding visibility graph. We use the method of successive difference for bin count: $y(\eta_i) = dn(\eta_{i+1})$ - $dn(\eta_i)$, where $dn(\eta_{i+1})$, $dn(\eta_i)$ are the event normalized pseudorapidity yields in two successive bins and projected it onto positive $(y, \eta)$ plane to get modified pseudorapidity distribution as required by the method of visibility graph [19]. Figure 3 represents the single event pseudorapidity distributions (a, c, e) and modified pseudorapidity distributions (b, d, f) in -2.5 < $\eta$ < 2.5 for pp collisions at $\sqrt{s}$ = 7 & 13 TeV and p-Pb collisions at $\sqrt{s_{NN}}$ = 5.02 TeV respectively for an arbitrarily chosen event.

The produced modified pseudorapidity distributions for the whole event sample are then mapped into the corresponding visibility graphs.
There are two types of processes in particle production in high energy collisions, i.e., soft processes and hard processes. With the increase of particle transverse momentum, a gradual transition from soft processes to hard processes occurs. To study the variation in fractality or complexity in particle production with the transverse momentum of produced particles, we have analysed four different thresholds for the charged particle transverse momentum ($p_{T,min}$): $p_T$ > 0.1, 0.3, 0.5, 0.7 GeV/c. The quantity degree denoted by $k$ is the number of connections that one particular node possesses with the other nodes present in the visibility graph of a single event. An overall degree distribution $P(k)$ for the entire event sample is obtained after combining the single event distributions. $P(k)$ vs $k$ graphs for the chosen energies are shown in Fig. 4 for four $p_T$ thresholds as mentioned above indicating the transition from soft to hard processes. We have chosen the $p_T$ thresholds for close differences to see the variation of overall degree distribution $P(k)$ for a smooth transition in particle production mechanisms. For better observability, the $P(k)$ values are added by multiple of a constant ($\alpha$ = 0.01) for different $p_T$ thresholds as specified in the legend of Fig. 4. It is evident from Fig. 4 that the power-law relationship of Eq. 2 is satisfied for all the energies and all $p_T$ thresholds. This result provides evidence in favour of the fractal nature of fluctuation patterns in multiparticle production in small collision systems in different centre-of-mass energies.
To extract the value of PSVG, we have fitted each degree distribution [$P(k)$ vs $k$] with a nonlinear function following Eq. 2 using the MINUIT program in the ROOT analysis framework [65] and the $\chi^2$ per degrees of freedom of each fit suggests acceptable goodness of fit of the graphs. Figure 4 shows the fitting exercise for all the $p_T$ thresholds in all three collision energies for EPOS3 generated pp and p-Pb events. The topological PSVG parameter, $\lambda_p$, for different $p_T$ thresholds are extracted from the fitting of each graph and are tabulated in Table 1.
Figure 5 shows the variation of the PSVG factor, $\lambda_p$, with four $p_T$ thresholds for EPOS3 generated ppevents at $\sqrt{s}$ = 7 & 13 TeV and p-Pb events at $\sqrt{s_{NN}}$ = 5.02 TeV. We observe that the amount of complexity or fractality increases gradually from low to high $p_T$ thresholds and the variation is almost similar for both the energies in pp collisions. In the case of p-Pb events, the increase of $\lambda_p$ is not very sharp for the same $p_T$ threshold values.

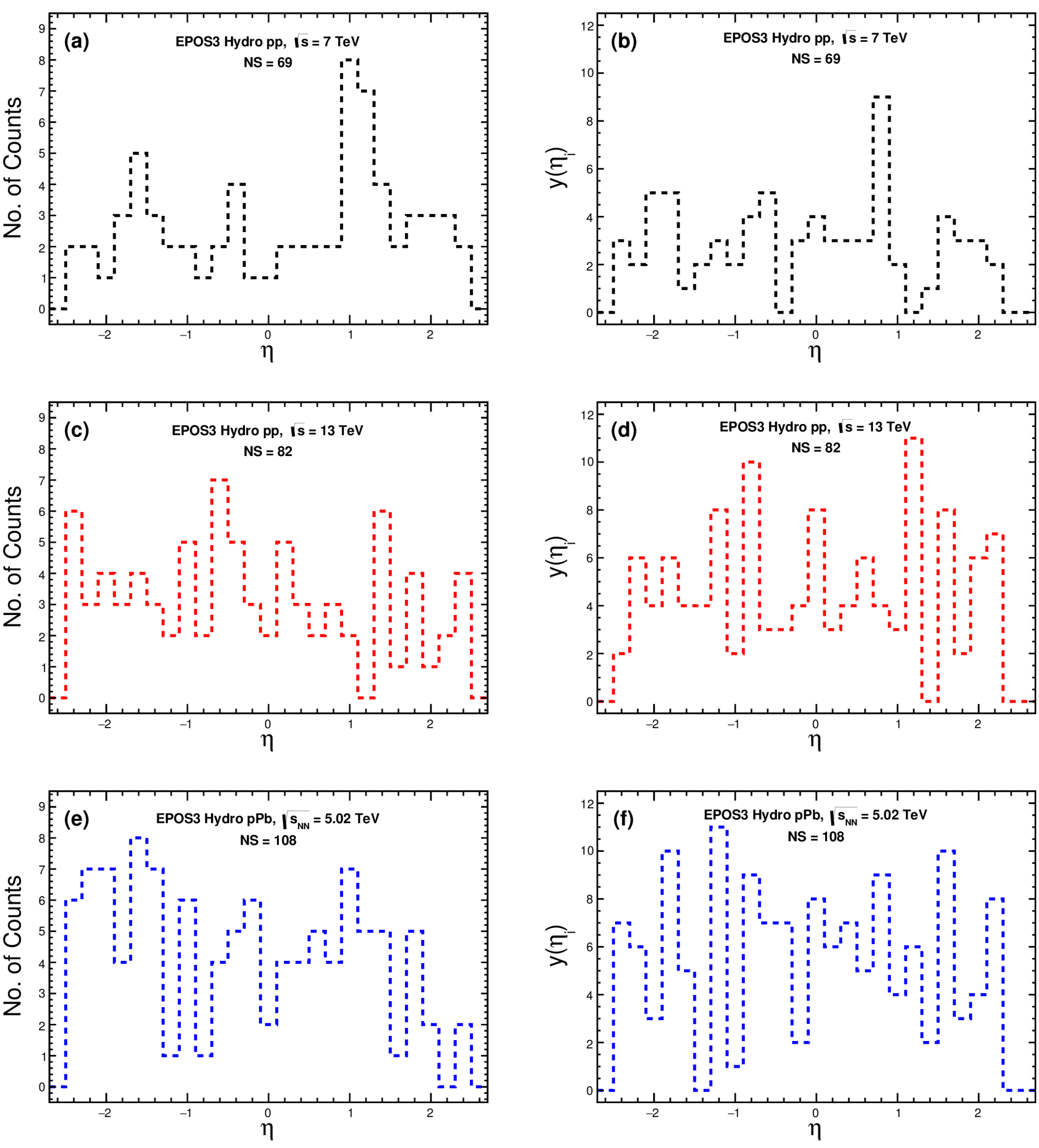

Fig. 3: (Color online) The single event (a, c, e) pseudorapidity distribution and (b, d, f) modified pseudorapidity distribution for EPOS3 generated pp and p-Pb events at $\sqrt{s}$ = 7 & 13 TeV and $\sqrt{s_{NN}}$ = 5.02 TeV respectively. NS is the charged particle multiplicity of an arbitrarily chosen event.

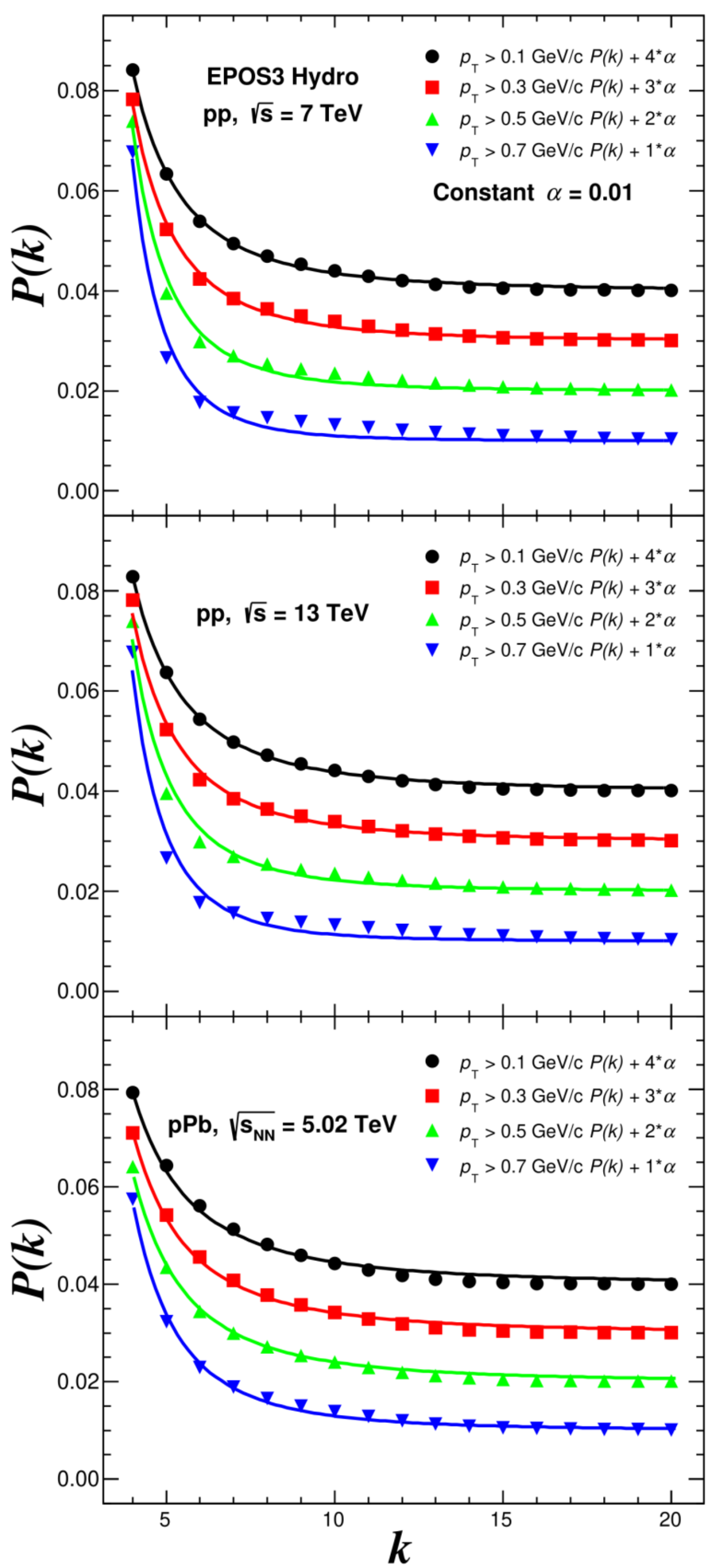

Fig. 4: (Color online) Degree distribution of visibility graph generated from EPOS3 simulated events in pp collisions at $\sqrt{s}$ = 7 TeV (top) & 13 TeV (middle) and in p-Pb collisions at $\sqrt{s_{NN}}$ = 5.02 TeV (bottom) for different transverse momentum thresholds ($p_{T,min}$).

| Collision Systems (Centre-of-Mass Energy) | $p_{T,min}$ (GeV/c) | $\lambda_p$ | $\chi^2$/NDF |
|---|---|---|---|
| **pp ($\sqrt{s}$ = 7 TeV)** | 0.1 | 2.752 ± 0.046 | 2.605 x $10^{-7}$ |
| | 0.3 | 3.080 ± 0.061 | 5.955 x $10^{-7}$ |
| | 0.5 | 3.685 ± 0.146 | 2.073 x $10^{-6}$ |
| | 0.7 | 4.363 ± 0.147 | 3.537 x $10^{-6}$ |
| **pp ($\sqrt{s}$ = 13 TeV)** | 0.1 | 2.642 ± 0.045 | 3.155 x $10^{-7}$ |
| | 0.3 | 2.879 ± 0.043 | 3.891 x $10^{-7}$ |
| | 0.5 | 3.393 ± 0.088 | 1.323 x $10^{-6}$ |
| | 0.7 | 4.045 ± 0.125 | 2.576 x $10^{-6}$ |
| **p-Pb ($\sqrt{s_{NN}}$ = 5.02 TeV)** | 0.1 | 2.363 ± 0.066 | 1.134 x $10^{-6}$ |
| | 0.3 | 2.514 ± 0.055 | 6.732 x $10^{-7}$ |
| | 0.5 | 2.586 ± 0.049 | 5.411 x $10^{-7}$ |
| | 0.7 | 3.018 ± 0.067 | 4.487 x $10^{-7}$ |

Table 1: Values of PSVG parameter and $\chi^2$/NDF for different $p_{T,min}$ values for EPOS3 simulated events in different collision systems.

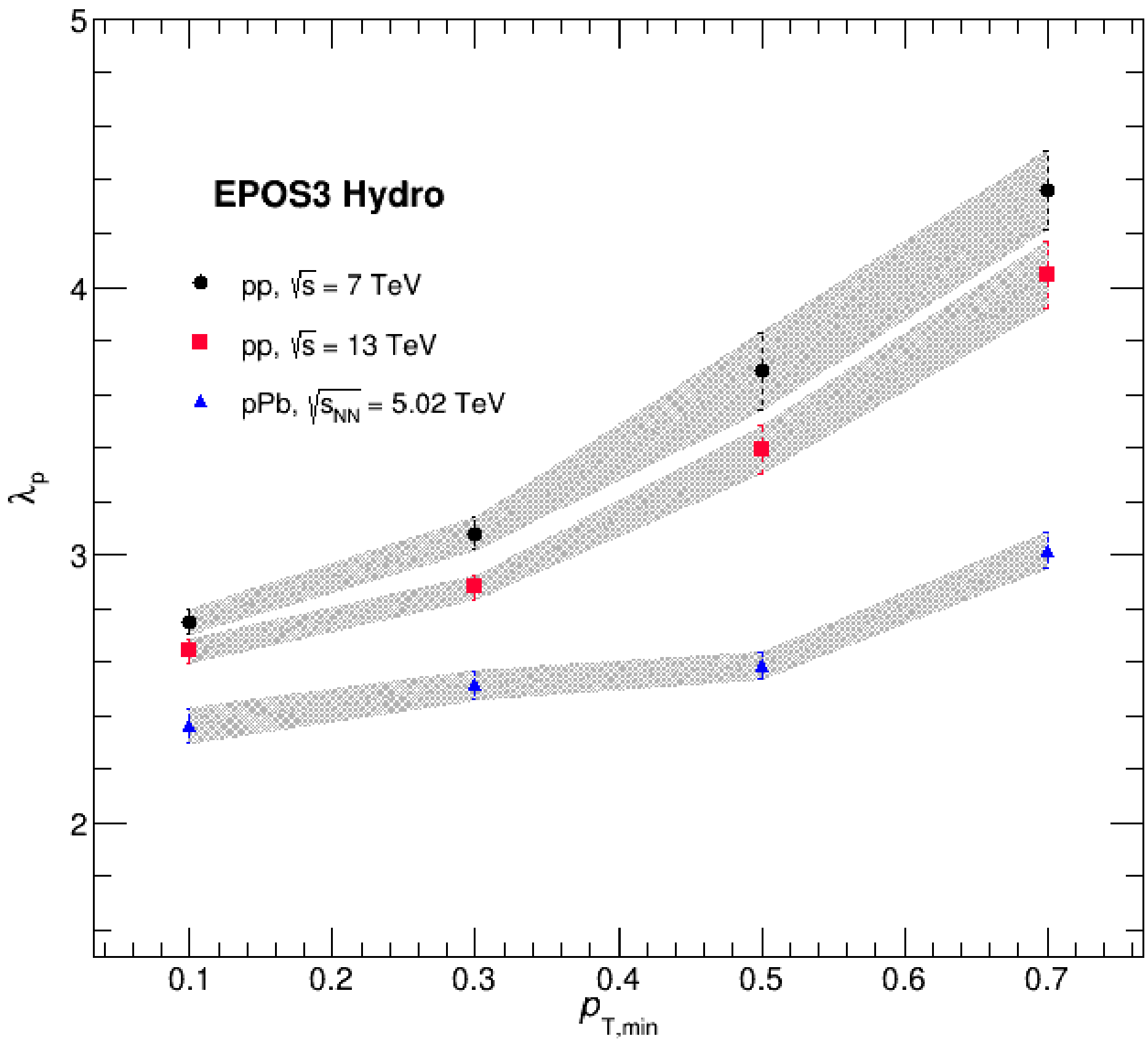

Fig. 5: (Color online) Variation of $\lambda_p$ with the minimum transverse momentum ($p_{T,min}$) for EPOS3 generated pp events at $\sqrt{s}$ = 7 & 13 TeV and p-Pb events at $\sqrt{s_{NN}}$ = 5.02 TeV. The error bars represent the fitting errors.

## 5. Conclusions

In this paper, the method of visibility graph-based nonlinear technique has been utilized to analyse fluctuation in high energy interactions at the LHC energies using a hybrid Monte-Carlo event generator EPOS3. The study reveals that:

- The event-by-event fluctuation of produced charged particles is fractal in nature for EPOS3 stimulated with hydro events for both pp and p-Pb collisions.
- The PSVG factor increases with the increase of minimum transverse momentum, $p_{T,min}$ in pp collisions for both the chosen centre-of-mass energies. It indicates that with the gradual transition from soft processes to hard processes, particle production becomes more complex in nature. In the case of p-Pb events, the PSVG factor does not show a sizeable increase with the increasing $p_{T,min}$ for the chosen $p_T$ thresholds. This could be due to the fact that with larger systems, the transition from soft to hard processes may be more prominent at higher $p_T$ compared to our chosen $p_T$ range.
- We also observed a clear system size dependency (pp and p-Pb) of the PSVG parameter for each of the $p_T$ thresholds which indicates that with larger colliding systems the complexity or the fractality in multiparticle production in high energy collisions decreases [66].

- In the case of pp collisions, the PSVG parameter increases with the decrease of the colliding centre-of-mass energy for each minimum transverse momentum, $p_{T,min}$. It also indicates that the complexity or fractal geometry of the multi-particle production mechanism decreases with the increase of energy of colliding beams for a fixed $p_T$ threshold value [66].

In this paper, we have presented an in-depth study on the fractal property of fluctuation patterns using the EPOS3 simulated pp and p-Pb events at $\sqrt{s}$ = 7 & 13 TeV and at $\sqrt{s_{NN}}$ = 5.02 TeV respectively with hydrodynamical evolution of produced particles for the first time. A comparative study with the LHC data will offer a check of the robustness of the EPOS3 model in terms of a new dimension, fractality.

## Conflicts of Interest

The authors declare that there is no conflict of interest regarding the publication of this article.

## Acknowledgments

The authors are grateful to Prof. Dr. Klaus Werner for providing them with the EPOS3 code. The authors are grateful to the Grid Computing Team of VECC, Kolkata and the Cluster Computing Team of the Department of Physics, Jadavpur University (JU) for providing an uninterrupted facility for event generation and analyses. We also gratefully acknowledge the financial support from the DST-INDIA under the scheme of the mega facilities for basic research and DHESTBT, WB. One of the authors (S. K.) acknowledges the financial support from UGC-INDIA Dr D. S. Kothari Post-Doctoral Fellowship under grant No. F.42/2006(BSR)/PH/19-20/0039.